\documentclass[aps,prc,twocolumn,groupedaddress,showpacs,amsmath,amssymb,floatfix,superscriptaddress]{revtex4}
\usepackage{multirow}
\usepackage{graphicx}
\usepackage{times}
\usepackage{color}
\newcommand{\PerTonDay}{(ton$\cdot$day)$^{-1}$}

\begin{document}

\bibliographystyle{h-physrev3}

\title{ Search for Majorana Neutrinos near the Inverted Mass Hierarchy Region with KamLAND-Zen }

\newcommand{\tohoku}{\affiliation{Research Center for Neutrino
    Science, Tohoku University, Sendai 980-8578, Japan}}
\newcommand{\osaka}{\affiliation{Graduate School of 
    Science, Osaka University, Toyonaka, Osaka 560-0043, Japan}}
\newcommand{\tokushima}{\affiliation{Faculty of Integrated Arts and Science, 
    University of Tokushima, Tokushima, 770-8502, Japan}}
\newcommand{\lbl}{\affiliation{Physics Department, University of
    California, Berkeley, and \\ Lawrence Berkeley National Laboratory, 
    Berkeley, California 94720, USA}}
\newcommand{\mitech}{\affiliation{Massachusetts Institute of Technology, 
    Cambridge, Massachusetts 02139, USA}}
\newcommand{\ut}{\affiliation{Department of Physics and
    Astronomy, University of Tennessee, Knoxville, Tennessee 37996, USA}}
\newcommand{\tunl}{\affiliation{Triangle Universities Nuclear
    Laboratory, Durham, North Carolina 27708, USA and \\
    Physics Departments at Duke University, North Carolina Central University,
    and the University of North Carolina at Chapel Hill}}
\newcommand{\ipmu}{\affiliation{Kavli Institute for the Physics and Mathematics of the Universe (WPI), 
    The University of Tokyo Institutes for Advanced Study, 
    The University of Tokyo, Kashiwa, Chiba 277-8583, Japan}}
\newcommand{\nikhef}{\affiliation{Nikhef and the University of Amsterdam, 
    Science Park, Amsterdam, the Netherlands}}
\newcommand{\washington}{\affiliation{Center for Experimental Nuclear Physics and Astrophysics, 
    University of Washington, Seattle, Washington 98195, USA}}

%
%
\author{A.~Gando}\tohoku
\author{Y.~Gando}\tohoku
\author{T.~Hachiya}\tohoku
\author{A.~Hayashi}\tohoku
\author{S.~Hayashida}\tohoku
\author{H.~Ikeda}\tohoku
\author{K.~Inoue}\tohoku\ipmu
\author{K.~Ishidoshiro}\tohoku
\author{Y.~Karino}\tohoku
\author{M.~Koga}\tohoku\ipmu
\author{S.~Matsuda}\tohoku
\author{T.~Mitsui}\tohoku
\author{K.~Nakamura}\tohoku\ipmu
\author{ S.~Obara}\tohoku
\author{T.~Oura}\tohoku
\author{H.~Ozaki}\tohoku
\author{I.~Shimizu}\tohoku
\author{Y.~Shirahata}\tohoku
\author{J.~Shirai}\tohoku
\author{A.~Suzuki}\tohoku
\author{T.~Takai}\tohoku
\author{K.~Tamae}\tohoku
\author{Y.~Teraoka}\tohoku
\author{K.~Ueshima}\tohoku
\author{H.~Watanabe}\tohoku

\author{A.~Kozlov}\ipmu
\author{Y.~Takemoto}\ipmu

\author{S.~Yoshida}\osaka

\author{K.~Fushimi}\tokushima

\author{T.I.~Banks}\lbl
\author{B.E.~Berger}\ipmu\lbl
\author{B.K.~Fujikawa}\ipmu\lbl
\author{T.~O'Donnell}\lbl

\author{L.A.~Winslow}\mitech

\author{Y.~Efremenko}\ipmu\ut

\author{H.J.~Karwowski}\tunl
\author{D.M.~Markoff}\tunl
\author{W.~Tornow}\ipmu\tunl

\author{J.A.~Detwiler}\ipmu\washington
\author{S.~Enomoto}\ipmu\washington

\author{M.P.~Decowski}\ipmu\nikhef

\collaboration{KamLAND-Zen Collaboration}\noaffiliation

\date{\today}

\begin{abstract}
We present an improved search for neutrinoless double-beta ($0\nu\beta\beta$) decay of $^{136}$Xe in the \mbox{KamLAND-Zen} experiment. Owing to purification of the xenon-loaded liquid scintillator, we achieved a significant reduction of the $^{110m}$Ag contaminant identified in previous searches. Combining the results from the first and second phase, we obtain a lower limit for the $0\nu\beta\beta$ decay half-life of $T_{1/2}^{0\nu} > 1.07 \times 10^{26}$\,yr at 90\% C.L., an almost sixfold improvement over previous limits. Using commonly adopted nuclear matrix element calculations, the corresponding upper limits on the effective Majorana neutrino mass are in the range 61 -- 165\,meV. For the most optimistic nuclear matrix elements, this limit reaches the bottom of the quasi-degenerate neutrino mass region.

\end{abstract}


\maketitle

Neutrinoless double-beta ($0\nu\beta\beta$) decay is an exotic nuclear process predicted by extensions of the Standard Model of particle physics. Observation of this decay would demonstrate the non-conservation of lepton number, and prove that neutrinos have a Majorana mass component. In the framework of light Majorana neutrino exchange, its decay rate is proportional to the square of the effective Majorana neutrino mass $\left<m_{\beta\beta}\right> \equiv \left| \sum_{i} U_{ei}^{2} m_{\nu_{i}}\right|$. Recent $0\nu\beta\beta$ searches~\cite{DellOro2016} involving $^{76}$Ge (GERDA~\cite{Agostini2013}) and $^{136}$Xe (\mbox{KamLAND-Zen}~\cite{Gando2013a} and \mbox{EXO-200}~\cite{Albert2014b}) provide upper limits on $\left<m_{\beta\beta}\right>$ of $\sim$0.2 -- 0.4\,eV using available nuclear matrix element (NME) values from the literature. The sensitivities of these searches correspond to mass scales in the so-called quasi-degenerate mass region. 

\mbox{KamLAND-Zen} is a double-beta decay experiment which exploits the existing detection infrastructure and radio-purity of KamLAND~\cite{Gando2015,Gando2013b}. The \mbox{KamLAND-Zen} detector consists of \mbox{13\,tons} of Xe-loaded liquid scintillator~(\mbox{Xe-LS}) contained in a 3.08-m-diameter spherical inner balloon (IB) located at the center of the KamLAND detector. The IB is constructed from 25-$\mu$m-thick transparent nylon film and is surrounded by \mbox{1\,kton}  of liquid scintillator (LS) contained in a 13-m-diameter spherical outer balloon (OB). The outer LS acts as an active shield. The scintillation photons are viewed by 1,879 photomultiplier tubes (PMTs) mounted on the inner surface of the containment vessel. The \mbox{Xe-LS} consists of 80.7\% decane and 19.3\% pseudocumene (1,2,4-trimethylbenzene) by volume, 2.29\,g/liter of the fluor PPO (2,5-diphenyloxazole), and $(2.91 \pm 0.04)$\% by weight of isotopically enriched xenon gas. The isotopic abundances in the enriched xenon were measured by a residual gas analyzer to be $(90.77 \pm 0.08)\%$ \mbox{$^{136}$Xe}, $(8.96 \pm 0.02)\%$ \mbox{$^{134}$Xe}. Other xenon isotopes have negligible presence. The two electrons emitted from $^{136}$Xe $\beta\beta$ decay produce scintillation light and their summed energy is observed. Hypothetical $0\nu\beta\beta$ decays would produce a peak at the $Q$-value of the decay, distinguishable from $2\nu\beta\beta$ decays which have a continuous spectrum. 

In the first phase of KamLAND-Zen (\mbox{Phase-I})~\cite{Gando2013a}, we obtained a lower limit of \mbox{$T_{1/2}^{0\nu} > 1.9 \times 10^{25}$\,yr} (90\% C.L.) on the $^{136}$Xe $0\nu\beta\beta$  decay half-life. The sensitivity of the \mbox{Phase-I} search was limited by the presence of an unexpected background peak, consistent with $^{110m}$Ag $\beta^{-}$ decay (\mbox{$\tau=360$\,day}, \mbox{$Q = 3.01$\,MeV}), just above the 2.458 MeV $Q$-value of $^{136}$Xe $\beta\beta$ decay. After completing \mbox{Phase-I}, we embarked on a \mbox{Xe-LS} purification campaign that continued for 18 months. First we extracted the \mbox{Xe-LS} in small batches from the IB through a Teflon tube whose intake was near the bottom of the IB volume. We then isolated and stored the Xe before placing the Xe-depleted LS back to the top of the IB where it was later replaced by new LS. This new LS was initially purified by water extraction followed by vacuum distillation. The replacement of the Xe-depleted LS was performed in three cycles equivalent to one IB volume exchange for each cycle. The LS was purified by vacuum distillation during each cycle. We also purified a mix of recovered and new Xe through distillation and refining with a heated zirconium getter. Finally, the Xe was dissolved into the purified LS. In December 2013, we started the second science run (\mbox{Phase-II}), and found a reduction of $^{110m}$Ag by more than a factor of 10. We report on the analysis of the complete \mbox{Phase-II} data set, collected between December 11, 2013, and October 27, 2015. The total livetime is 534.5\,days after muon spallation cuts, discussed later. This corresponds to an exposure of 504\,kg-yr of $^{136}$Xe with the whole \mbox{Xe-LS} volume.

Following the end of \mbox{Phase-II}, we performed a detector calibration campaign using radioactive sources deployed at various positions along the central axis of the IB. The event position reconstruction --- determined from the scintillation photon arrival times --- reproduces the known source positions to within 2.0\,cm; the reconstruction performance is better than 1.0\,cm for events occurring within 1\,m of the IB center. The energy scale was studied using $\gamma$-rays from $^{60}$Co, $^{68}$Ge, and $^{137}$Cs radioactive sources, $\gamma$-rays from the capture of spallation neutrons on protons and $^{12}$C, and $\beta + \gamma$-ray emissions from $^{214}$Bi, a daughter of $^{222}$Rn ($\tau = 5.5\,{\rm day}$) that was introduced during the \mbox{Xe-LS} purification. The calibration data indicate that the reconstructed energy varies by less than 1.0\% throughout the \mbox{Xe-LS} volume, and the time variation of the energy scale is less than 1.0\%. Uncertainties from the nonlinear energy response due to scintillator quenching and Cherenkov light production are constrained by the calibrations. The light yield of the Xe-LS is 7\% lower than that of the outer LS, which is corrected in the detector simulation, while the non-linearities for both the LS regions are consistent. The observed energy resolution is $\sigma \sim 7.3\%/\sqrt{E{\rm (MeV)}}$, slightly worse relative to \mbox{Phase-I} due to an increased number of dead PMTs. 

We apply the following series of cuts to select $\beta\beta$ decay events: (i) The reconstructed vertex must be within 2.0\,m of the detector center. (ii) Muons and events within 2\,ms after muons are rejected. (iii) $^{214}$Bi-$^{214}$Po ($\tau$=237\,$\mu$s) decays are eliminated by a delayed coincidence tag, requiring the time and distance between the prompt $^{214}$Bi and delayed $^{214}$Po decay-events to be less than 1.9\,ms and 1.7\,m, respectively. The cut removes $(99.95 \pm 0.01)\%$ of $^{214}$Bi-$^{214}$Po decays, where the inefficiency is dominated by the timing cut, and the uncertainty is estimated from analysis of periods with high Rn levels. The same cut is not effective for $^{212}$Bi-$^{212}$Po ($\tau$=0.4\,$\mu$s) decays which occur within a single $\sim$200-ns-long data acquisition event window. Therefore, the cut is augmented with a double-pulse identification in the photon arrival time distribution after subtracting the time of flight from the vertex to each PMT. The $^{212}$Bi-$^{212}$Po rejection efficiency is $(95 \pm 3)\%$, confirmed with high-Rn data. (iv) Reactor $\overline{\nu}_{e}$'s identified by a delayed coincidence of positrons and neutron-capture $\gamma$'s~\cite{Gando2013b} are discarded. (v) Poorly reconstructed events are rejected. These events are tagged using a vertex-time-charge discriminator which measures how well the observed PMT time-charge distributions agree with those expected based on the reconstructed vertex~\cite{Abe2011}. The total cut inefficiency for $\beta\beta$ events is less than 0.1\%.

\begin{figure}[t]
\begin{center}
\includegraphics[width=1.0\columnwidth]{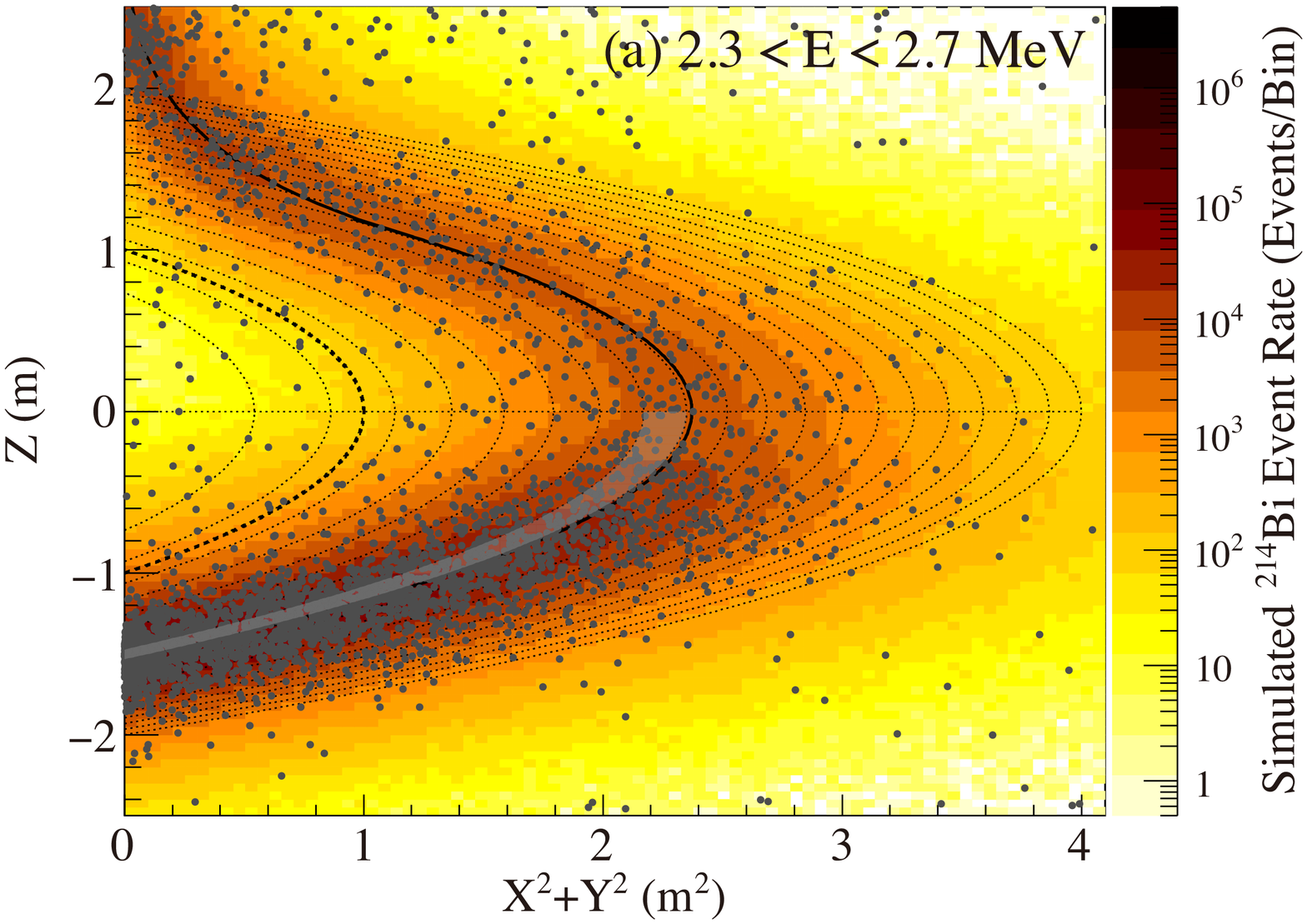}
\\
\vspace{0.1cm}
\includegraphics[width=1.0\columnwidth]{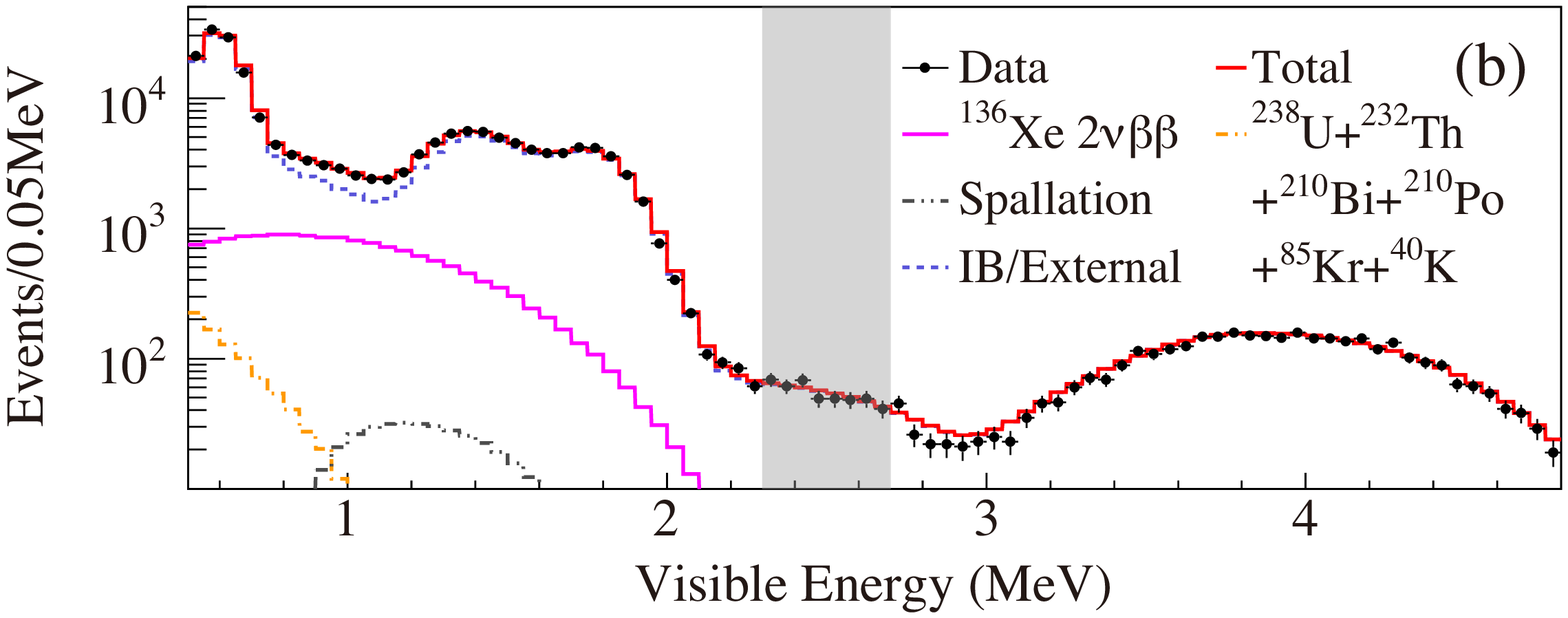}
\\
\vspace{0.2cm}
\includegraphics[width=1.0\columnwidth]{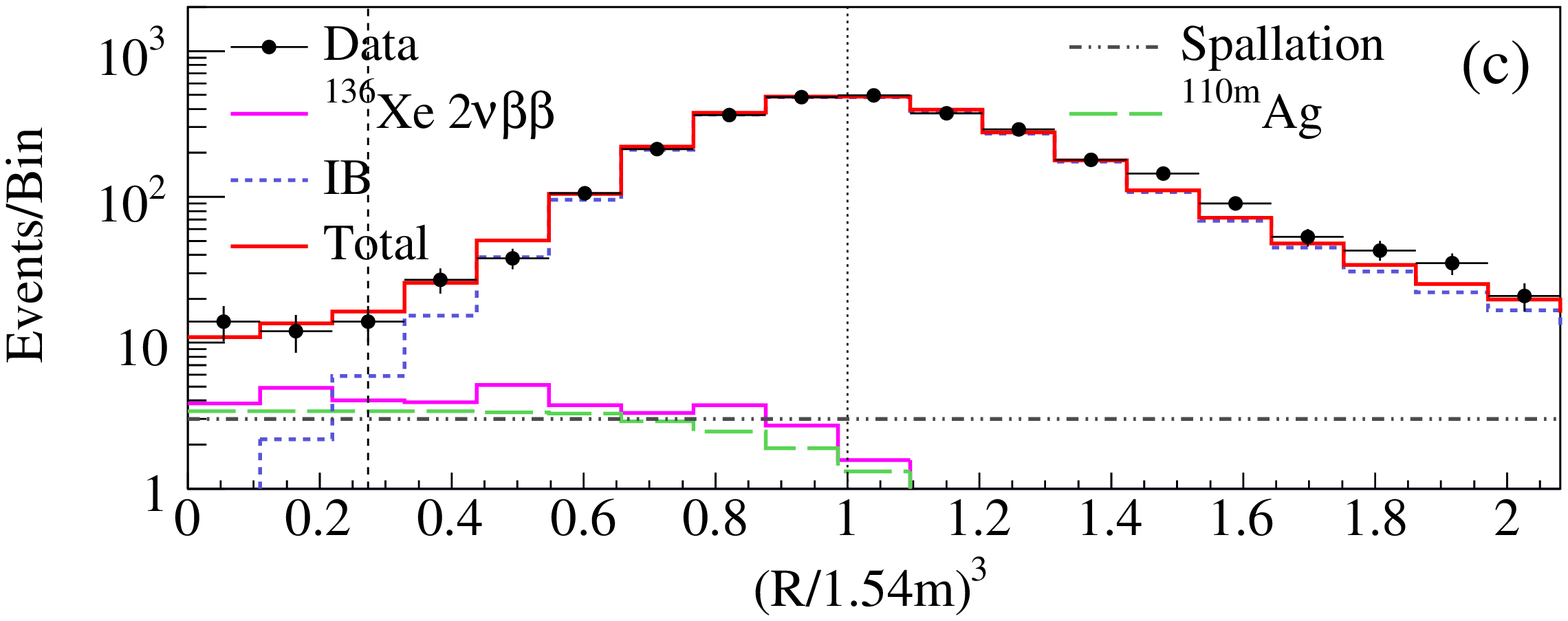}
\vspace{-0.9cm}
\end{center}
  \caption{(a) Vertex distribution of candidate events (black points) and reproduced $^{214}$Bi background events in a MC simulation (color histogram) for $2.3 < E < 2.7\,{\rm MeV}$ (the $0\nu\beta\beta$ window). The normalization of the MC event histogram is arbitrary. The solid and thick dashed lines indicate the shape of the IB and the 1-m-radius spherical volume, respectively. The thin dashed lines illustrate the shape of the equal-volume spherical half-shells which compose the 2-m-radius spherical fiducial volume for the $0\nu\beta\beta$ analysis. (b) An example of the energy spectrum in a volume bin with high $^{214}$Bi background events around the lower part of the IB film (shaded region in (a) at $1.47 < R < 1.53\,{\rm m}, z < 0$). (c) $R^{3}$ vertex distribution of candidate events in the $0\nu\beta\beta$ window. The curves show the best-fit background model components.}
  \vspace{-0.3cm}
  \label{figure:vertex}
\end{figure}

Background sources external to the \mbox{Xe-LS} are dominated by radioactive impurities on the IB film. Based on a spectral fit to events reconstructed around the IB, we find that the dominant background sources are $^{134}$Cs ($\beta + \gamma$, $\tau = 2.97\,{\rm yr}$) in the energy region $1.2 < E < 2.0\,{\rm MeV}$ ($2\nu\beta\beta$ window), and $^{214}$Bi in the region $2.3 < E < 2.7\,{\rm MeV}$ ($0\nu\beta\beta$ window). The observed activity ratio of $^{134}$Cs to $^{137}$Cs (0.662\,MeV $\gamma$, $\tau = 43.4\,{\rm yr}$) indicates that the IB film was contaminated by fallout from the Fukushima-I reactor accident in 2011~\cite{Gando2012a}. $^{214}$Bi is a daughter of $^{238}$U, a naturally occurring contaminant. The observed rate of $^{214}$Bi decays indicates that the $^{238}$U concentration in the nylon film is 0.16\,ppb assuming secular equilibrium, while the \mbox{{\it ex situ}} measurement by ICP-MS yielded 2\,ppt. The non-uniform $^{214}$Bi event distribution observed on the IB indicates that this discrepancy is caused by dust contamination rather than decay chain non-equilibrium. Figure~\ref{figure:vertex}~(a) shows the vertex distribution of candidate events, and the predicted $^{214}$Bi background events from a Monte Carlo (MC) simulation in the $0\nu\beta\beta$ window. The z-distribution of $^{214}$Bi decays on the IB is evaluated from the data, and used for the radioactive decay model in the MC. For the $^{214}$Bi background, the vertex dispersion model was constructed from a full MC simulation based on \texttt{Geant4}~\cite{Agostinelli2003,Allison2006} including decay-particle tracking, scintillation photon processes, and finite PMT timing resolution. This MC reproduces the observed vertex distance between $^{214}$Bi and $^{214}$Po sequential decay events from the initial $^{222}$Rn contamination within the \mbox{Xe-LS}.

\begin{table}[t]
\caption[]{Summary of the number of observed events, and the estimated and best-fit background contributions in the energy region $2.3 < E < 2.7\,{\rm MeV}$ ($0\nu\beta\beta$ window) within the 1-m-radius spherical volume for each of the two time periods.}
\label{table:background}

\begin{center}
\begin{tabular}{c c c c c}
\hline
\hline
 & \multicolumn{2}{c}{Period-1} & \multicolumn{2}{c}{Period-2} \\
 & \multicolumn{2}{c}{(270.7\,days)} & \multicolumn{2}{c}{(263.8\,days)} \\
\hline
Observed events & \multicolumn{2}{c}{22} & \multicolumn{2}{c}{11} \vspace{0.2cm} \\
Background & Estimated & Best-fit & Estimated & Best-fit \\
\hline
$^{136}$Xe $2\nu\beta\beta$ & - & 5.48 & - & 5.29 \\
\multicolumn{5}{c}{Residual radioactivity in \mbox{Xe-LS}} \\
\hline
$^{214}$Bi ($^{238}$U series) & $0.23 \pm 0.04$ & 0.25 & $0.028 \pm 0.005$ & 0.03 \\
$^{208}$Tl ($^{232}$Th series) & - & 0.001 & - & 0.001 \\
$^{110m}$Ag & - & 8.5 & - & 0.0 \\
\multicolumn{5}{c}{External (Radioactivity in IB)} \\
\hline
$^{214}$Bi ($^{238}$U series) & - & 2.56 & - & 2.45 \\
$^{208}$Tl ($^{232}$Th series) & - & 0.02 & - & 0.03 \\
$^{110m}$Ag & - & 0.003 & - & 0.002 \\
\multicolumn{5}{c}{Spallation products} \\
\hline
$^{10}$C & $2.7 \pm 0.7$ & 3.3 & $2.6 \pm 0.7$ & 2.8 \\
$^{6}$He & $0.07 \pm 0.18$ & 0.08 & $0.07 \pm 0.18$ & 0.08 \\
$^{12}$B & $0.15 \pm 0.04$ & 0.16 & $0.14 \pm 0.04$ & 0.15 \\
$^{137}$Xe & $0.5 \pm 0.2$ & 0.5 & $0.5 \pm 0.2$ & 0.4 \\
\hline
\hline
  \vspace{-0.9cm}
\end{tabular}
\end{center}
\end{table}

An enlarged 3.5-m-radius spherical volume was used to study a high statistics sample of muon spallation products and better constrain their background contributions. This included a region outside the IB. We assess a 22\% systematic uncertainty on the calculated spallation yields in the \mbox{Xe-LS}, taking account of the observed $(20 \pm 2)$\% increase in the spallation neutron flux in the \mbox{Xe-LS} relative to the outer LS. In the $0\nu\beta\beta$ window, events from $^{10}$C decays ($\beta^{+}$, \mbox{$\tau = 27.8$\,s}, \mbox{$Q = 3.65$\,MeV}) dominate the contribution from muon spallation. A triple-coincidence tag of a muon, a neutron identified by neutron-capture $\gamma$-rays, and the $^{10}$C decay~\cite{Abe2010}, reduces the $^{10}$C background with an efficiency of $(64 \pm 4)$\%. Post-muon spallation-neutron events are recorded by newly introduced dead-time free electronics. We apply spherical volume cuts ($\Delta R < 1.6\,{\rm m}$) around the reconstructed neutron vertices for 180\,s after the preceding muon. We estimate that the remaining $^{10}$C background after cuts is $(1.01 \pm 0.26) \times 10^{-2}$\,\PerTonDay, where ton is a unit of \mbox{Xe-LS} mass. Other shorter-lived products, e.g., $^{6}$He and $^{12}$B, are also reduced by the triple-coincidence tag and have a minor contribution to the background. The dead-time introduced by all the spallation cuts is 7\%. In the \mbox{Xe-LS}, long-lived $^{137}$Xe ($\beta^{-}$, \mbox{$\tau=5.5$\,min}, \mbox{$Q=4.17$\,MeV}) is a background source produced by neutron capture on $^{136}$Xe. Based on the spallation neutron rate and the $^{136}$Xe capture cross section~\cite{Albert2016}, the production yield of $^{137}$Xe is estimated to be $(3.9 \pm 2.0) \times 10^{-3}$\,\PerTonDay, which is consistent with the simulation study in \texttt{FLUKA}~\cite{Battistoni2007,Ferrari2005}.

We perform the $0\nu\beta\beta$ decay analysis using a 2-m-radius FV as described above to utilize the deployed $^{136}$Xe mass. However, the sensitivity is dominated by the innermost 1-m-radius spherical volume due to the background from the IB. The region outside this radius serves to strongly constrain the tails of the IB background extending into the innermost region. Further, anticipating the decay of the $^{110m}$Ag background identified in Phase-I, we divide the \mbox{Phase-II} data set into two equal time periods (\mbox{Period-1} and \mbox{Period-2}) each roughly equal to one average lifetime of the $^{110m}$Ag decay rate. Table~\ref{table:background} lists the number of observed events, and the estimated and best-fit background contributions in the $0\nu\beta\beta$ window within a 1-m-radius spherical volume for each of the two time periods. The fit is described in detail below. We find a precipitous decrease in the event rate in the $0\nu\beta\beta$ window in \mbox{Period-2}. The \mbox{Period-2} background components are well-constrained near the values listed in Table~\ref{table:background} with the exception of $^{110m}$Ag. The hypothesis of standard radioactive decay of the $0\nu\beta\beta$ window background with the decay rate of $^{110m}$Ag is disfavored relative to the hypothesis of a faster reduction at 96\% C.L. The origin of this apparent reduction of $^{110m}$Ag is unknown, but we speculate that much of it settled to the bottom of the IB where only a small fraction of $^{110m}$Ag decays are reconstructed in the inner \mbox{Xe-LS} volume. In order to allow the $0\nu\beta\beta$ window background the greatest freedom in the fit, the $0\nu\beta\beta$ decay analyses are performed independently for \mbox{Period-1} and \mbox{Period-2}.

\begin{figure}
\vspace{0.2cm}
\includegraphics[width=1.0\columnwidth]{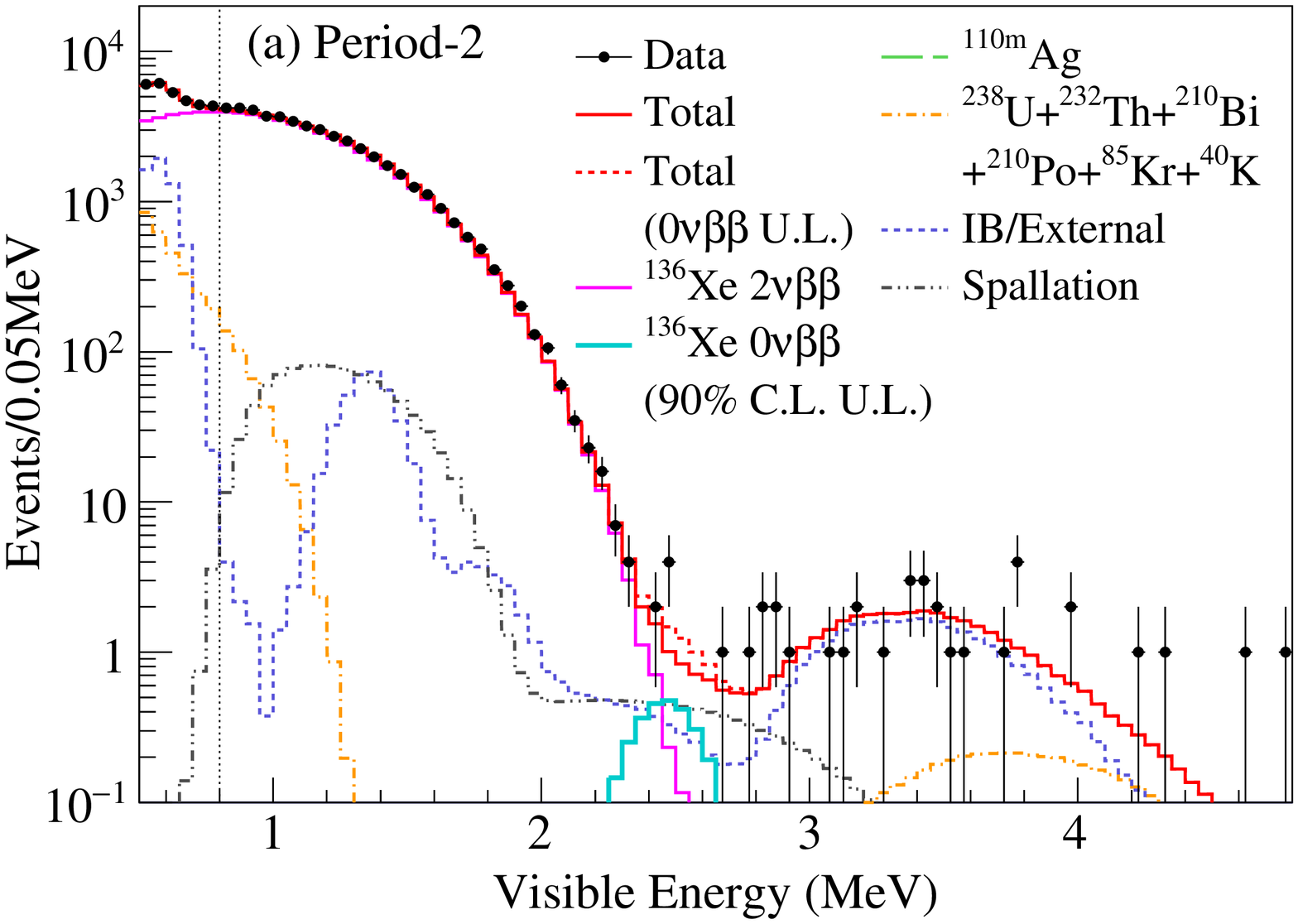}
\\
\vspace{0.3cm}
\includegraphics[width=1.0\columnwidth]{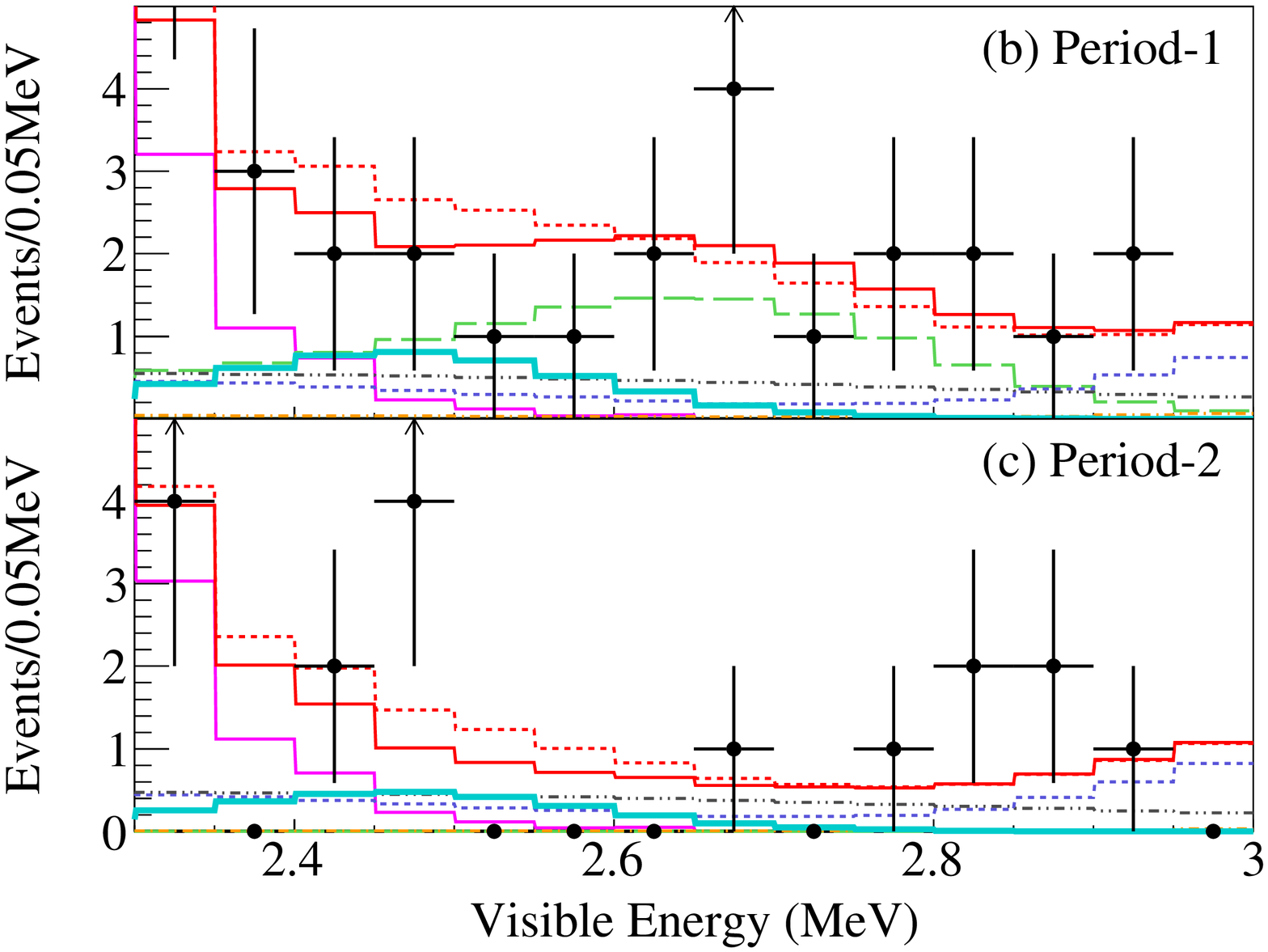}
\vspace{-0.5cm}
  \caption{(a) Energy spectrum of selected $\beta\beta$ candidates within a 1-m-radius spherical volume in \mbox{Period-2} drawn together with best-fit backgrounds, the $2\nu\beta\beta$ decay spectrum, and the 90\% C.L. upper limit for $0\nu\beta\beta$ decay. (b), (c) Closeup energy spectra for $2.3 < E < 3.0\,{\rm MeV}$ in \mbox{Period-1} and \mbox{Period-2}, respectively.}
  \vspace{-0.3cm}
  \label{figure:energy}
\end{figure}

The $2\nu\beta\beta$ decay rate can in principle be estimated from the same analysis used to derive the $0\nu\beta\beta$ decay limits, but the region outside of 1-m radius contributes negligibly to the $2\nu\beta\beta$ decay rate estimate and is dominated by systematic uncertainty arising from the IB background. To obtain a $2\nu\beta\beta$ decay rate free of such systematic uncertainty, we perform a separate estimate using a likelihood fit to the binned energy spectrum of the selected candidates between 0.5 MeV and 4.8 MeV, limited to a volume within the 1-m-radius spherical fiducial volume (FV$_{2\nu}$). The corresponding fiducial exposure of $^{136}$Xe is 126\,kg-yr. The contributions from major backgrounds in the \mbox{Xe-LS}, such as $^{85}$Kr, $^{40}$K, $^{210}$Bi, and the $^{228}$Th-$^{208}$Pb sub-chain of the $^{232}$Th series are free parameters and are left unconstrained in the spectral fit. We confirmed that the $^{134}$Cs contribution in the Xe-LS is negligible from a fit. The contributions from the $^{222}$Rn-$^{210}$Pb sub-chain of the $^{238}$U series, muon spallation products, and detector energy response model parameters are allowed to vary but are constrained by their independent estimations. The $2\nu\beta\beta$ decay rates for \mbox{Period-1} and \mbox{Period-2} are $100.1^{+1.1}_{-1.8}$\,(ton$\cdot$day)$^{-1}$ and $100.1^{+1.0}_{-0.9}$\,(ton$\cdot$day)$^{-1}$, respectively, and are in agreement within the statistical uncertainties. The resolution tail in $2\nu\beta\beta$ decays is an important background in the $0\nu\beta\beta$ analysis. Such tail events are reproduced in $^{214}$Bi decays with high-Rn data assuming the Gaussian resolution, indicating that a contribution from energy reconstruction failures is negligible.

We assess the systematic uncertainty of the FV$_{2\nu}$ cut based on the study of uniformly distributed $^{214}$Bi events from initial $^{222}$Rn contamination throughout the \mbox{Xe-LS}. We obtain a 3.0\% systematic error on FV$_{2\nu}$, consistent with the 1.0\,cm radial-vertex-bias in the source calibration data. Other sources of systematic uncertainty such as xenon mass (0.8\%), detector energy scale (0.3\%) and efficiency (0.2\%), and $^{136}$Xe enrichment (0.09\%), only have a small contribution; the overall uncertainty is 3.1\%. The measured $2\nu\beta\beta$ decay half-life of $^{136}$Xe is $T_{1/2}^{2\nu} = 2.21 \pm 0.02({\rm stat}) \pm 0.07({\rm syst}) \times 10^{21}$\,yr. This result is consistent with our previous result based on \mbox{Phase-I} data, $T_{1/2}^{2\nu} = 2.30 \pm 0.02({\rm stat}) \pm 0.12({\rm syst}) \times 10^{21}$\,yr~\cite{Gando2012b}, and with the result obtained by \mbox{EXO-200}, $T_{1/2}^{2\nu} = 2.165 \pm 0.016({\rm stat}) \pm 0.059({\rm syst}) \times 10^{21}$\,yr~\cite{Albert2014a}.

For the $0\nu\beta\beta$ analysis, using the larger 2-m-radius FV, the dominant $^{214}$Bi background on the IB is radially attenuated but larger in the lower hemisphere.  So we divide the FV into 20-equal-volume bins for each of the upper and lower hemispheres (see Fig.~\ref{figure:vertex}~(a)). We perform a simultaneous fit to the energy spectra for all volume bins. The z-dependence of $^{214}$Bi on the IB film is extracted from a fixed energy window dominated by these events. The $^{214}$Bi background contribution is then broken into two independent distributions in the upper and lower hemispheres whose normalizations are floated as free parameters. The fit reproduces the energy spectra for each volume bin; Fig.~\ref{figure:vertex}~(b) shows an example of the energy spectrum in a volume bin with high $^{214}$Bi background events around the IB film. The radial dependences of candidate events and best-fit background contributions in the $0\nu\beta\beta$ window are illustrated in Fig.~\ref{figure:vertex}~(c). The possible background contributions from $^{110m}$Ag are free parameters in the fit. We consider three independent components: $^{110m}$Ag uniformly dispersed in the \mbox{\mbox{Xe-LS}} volume, and on the surfaces of each the lower and upper IB films. We also examined non-uniform $^{110m}$Ag sources, with different assumed radial dependences, in the \mbox{Xe-LS} but determined that this has little impact on the $0\nu\beta\beta$ limit.

As described above, the fits are performed independently for \mbox{Period-1} and \mbox{Period-2} in the region $0.8 < E < 4.8\,{\rm MeV}$. We found no event excess over the background expectation for both data sets. The 90\% C.L. upper limits on the $^{136}$Xe $0\nu\beta\beta$ decay rate are $<$5.5\,(kton$\cdot$day)$^{-1}$ and $<$3.4\,(kton$\cdot$day)$^{-1}$ for \mbox{Period-1} and \mbox{Period-2}, respectively. To demonstrate the low background levels achieved in the $0\nu\beta\beta$ region, Fig.~\ref{figure:energy} shows the energy spectra within a 1-m-radius, together with the best-fit background composition and the 90\% C.L. upper limit for $0\nu\beta\beta$ decays. Combining the results, we obtain a 90\% C.L. upper limit of $<$2.4\,(kton$\cdot$day)$^{-1}$, or $T_{1/2}^{0\nu} > 9.2 \times 10^{25}$\,yr (90\% C.L.). We find a fit including potential backgrounds from $^{88}$Y, $^{208}$Bi, and $^{60}$Co~\cite{Gando2013a} does not change the obtained limit. A MC of an ensemble of experiments assuming the best-fit background spectrum without a $0\nu\beta\beta$ signal indicates a sensitivity of $5.6 \times 10^{25}$\,yr, and the probability of obtaining a limit stronger than the presented result is 12\%. For comparison, the sensitivity of an analysis in which the $^{110m}$Ag background rates in \mbox{Period-1} and \mbox{Period-2} are constrained to the $^{110m}$Ag half-life is $4.5 \times 10^{25}$\,yr.

Combining the \mbox{Phase-I} and \mbox{Phase-II} results, we obtain $T_{1/2}^{0\nu} > 1.07 \times 10^{26}$\,yr (90\% C.L.). This corresponds to an almost sixfold improvement over the previous \mbox{KamLAND-Zen} limit using only the \mbox{Phase-I} data, owing to a significant reduction of the $^{110m}$Ag contaminant and the increase in the exposure of $^{136}$Xe.

\begin{figure}
\includegraphics[width=1.0\columnwidth]{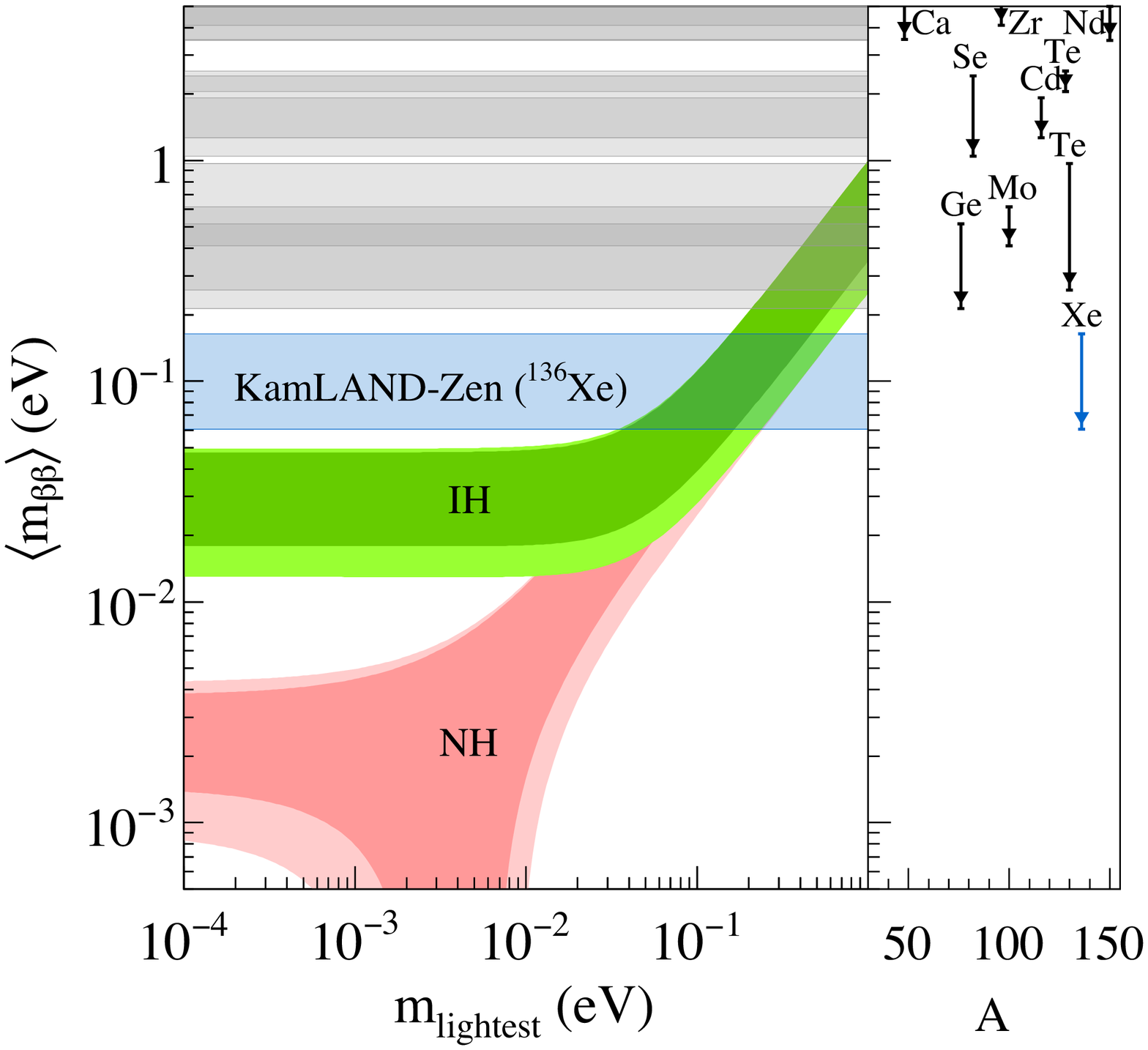}
\vspace{-0.7cm}
  \caption{Effective Majorana neutrino mass $\left<m_{\beta\beta}\right>$ as a function of the lightest neutrino mass $m_{\rm lightest}$. The dark shaded regions are the predictions based on best-fit values of neutrino oscillation parameters for the normal hierarchy (NH) and the inverted hierarchy (IH), and the light shaded regions indicate the $3\sigma$ ranges calculated from the oscillation parameter uncertainties~\cite{DellOro2014,Capozzi2014}. The horizontal bands indicate 90\% C.L. upper limits on $\left<m_{\beta\beta}\right>$ with $^{136}$Xe from \mbox{KamLAND-Zen} (this work), and with other nuclei from Ref.~\cite{Agostini2013,Alfonso2015,Arnold2015,Barabash2011b}, considering an improved phase space factor calculation~\cite{Kotila2012,Stoica2013} and commonly used NME calculations~\cite{Rodriguez2010,Menendez2009,Barea2015,Hyvarinen2015,Meroni2013,Simkovic2013,Mustonen2013}. The side-panel shows the corresponding limits for each nucleus as a function of the mass number.}
  \vspace{-0.3cm}
  \label{figure:effective_mass}
\end{figure}

From the limit on the $^{136}$Xe $0\nu\beta\beta$ decay half-life, we obtain a 90\% C.L. upper limit of $\left<m_{\beta\beta}\right> < (61 \text{ -- } 165)\,{\rm meV}$ using an improved phase space factor calculation~\cite{Kotila2012,Stoica2013} and commonly used NME calculations~\cite{Rodriguez2010,Menendez2009,Barea2015,Hyvarinen2015,Meroni2013,Simkovic2013,Mustonen2013} assuming the axial coupling constant $g_{A} \simeq 1.27$. Figure~\ref{figure:effective_mass} illustrates the allowed range of $\left<m_{\beta\beta}\right>$ as a function of the lightest neutrino mass $m_{\rm lightest}$ under the assumption that the decay mechanism is dominated by exchange of a pure-Majorana Standard Model neutrino. The shaded regions include the uncertainties in $U_{ei}$ and the neutrino mass splitting, for each hierarchy. Also drawn are the experimental limits from the $0\nu\beta\beta$ decay searches for each nucleus~\cite{Agostini2013,Alfonso2015,Arnold2015,Barabash2011b}. The upper limit on $\left<m_{\beta\beta}\right>$ from \mbox{KamLAND-Zen} is the most stringent, and it also provides the strongest constraint on $m_{\rm lightest}$ considering extreme cases of the combination of CP phases and the uncertainties from neutrino oscillation parameters~\cite{DellOro2014,Capozzi2014}. We obtain a 90\% C.L. upper limit of $m_{\rm lightest} < (180 - 480)\,{\rm meV}$.

In conclusion, we have demonstrated effective background reduction in the Xe-loaded liquid scintillator by purification, and enhanced the $0\nu\beta\beta$ decay search sensitivity in \mbox{KamLAND-Zen}. Our search constrains the mass scale to lie below $\sim$100\,meV, and the most advantageous nuclear matrix element calculations indicate an effective Majorana neutrino mass limit near the bottom of the quasi-degenerate neutrino mass region. The current \mbox{KamLAND-Zen} search is limited by backgrounds from $^{214}$Bi, $^{110m}$Ag, muon spallation and partially by the tail of $2\nu\beta\beta$ decays. In order to improve the search sensitivity, we plan to upgrade the \mbox{KamLAND-Zen} experiment with a larger \mbox{Xe-LS} volume loaded with 800\,kg of enriched Xe, corresponding to a twofold increase in $^{136}$Xe, contained in a larger balloon with lower radioactive background contaminants. If further radioactive background reduction is achieved, the background will be dominated by muon spallation, which can be further reduced by optimization of the spallation cut criteria. Such an improved search will allow $\left<m_{\beta\beta}\right>$ to be probed below 50\,meV, starting to constrain the inverted mass hierarchy region under the assumption that neutrinos are Majorana particles. The sensitivity of the experiment can be pushed further by improving the energy resolution to minimize the leakage of the $2\nu\beta\beta$ tail into the $0\nu\beta\beta$ analysis window. Such improvement is the target of a future detector upgrade.

The authors wish to acknowledge Prof.~A.~Piepke for providing radioactive sources for KamLAND. The \mbox{KamLAND-Zen} experiment is supported by JSPS KAKENHI Grant Numbers 21000001 and 26104002; the World Premier International Research Center Initiative (WPI Initiative), MEXT, Japan; Stichting FOM in the Netherlands; and under the U.S. Department of Energy (DOE) Grant No.\,DE-AC02-05CH11231, as well as other DOE and NSF grants to individual institutions. The Kamioka Mining and Smelting Company has provided service for activities in the mine. We acknowledge the support of NII for SINET4.

\bibliography{DoubleBeta}

\end{document}